\documentclass{iau}
\usepackage{graphicx}

\newcommand{\chisq}{$\chi^{2}$}
\newcommand{\G}{G308.3-1.4}

\title[IAUS291.~~G308.3-1.4 and its CCO in X-rays] 
{X-ray properties of G308.3-1.4 and its central compact object} 

\author[K. A. Seo et al.]  
{ K. A. Seo$^1$, C. Y. Hui$^1$, R. H. H. Huang$^2$,
  L. Trepl$^3$, T.-N. Lu$^2$,\\
A. K. H. Kong$^2$\thanks{Golden Jade Fellow of Kenda Foundation, Taiwan}
 \and F. M. Walter$^4$}

\affiliation{$^1$Department of Astronomy and Space Science, 
Chungnam National University, \\Daejeon 305-764, Korea \\[\affilskip]

$^2$Institute of Astronomy and Department of Physics, National 
Tsing Hua University, \\Hsinchu, Taiwan \\[\affilskip]

$^3$Astrophysikalisches Institut und Universit\"{a}ts-Sternwarte,
Universit\"{a}t Jena, \\Schillerg\"{a}{\ss}chen 2-3, 07745 Jena,
Germany\\ [\affilskip]

$^4$Department of Physics and Astronomy, Stony Brook University, Stony
Brook, NY 11794-3800, USA \\ [\affilskip]
}

\pubyear{2012}
\volume{291}  
\jname{\mbox{Neutron Stars and Pulsars: Challenges and Opportunities after 80 years}}
\editors{J. van Leeuwen, ed.} 
\begin{document}

\maketitle
\begin{abstract}
We present a short \emph{Chandra} observation that confirms a previous unidentified extended X-ray source, 
G308.3-1.4, as a new supernova remnant (SNR) in the 
Milky Way. Apart from identifying its SNR nature, a bright X-ray point source has also been discovered at 
the geometrical center. Its X-ray spectral properties are similar to those of a particular class of neutron star 
known as central compact objects (CCOs). On the other hand, the
optical properties of this counterpart suggests 
it to be a late-type star. Together with the interesting $\sim1.4$~hours X-ray periodicity found by \emph{Chandra}, this 
system can possibly provide the first direct evidence of a compact binary survived in a supernova explosion.
\end{abstract}



              
\section{Introduction}
Recently, we initiated an extensive identification campaign of
unidentified extended ROSAT All-Sky Survey (RASS) objects (Hui et
al. 2012). The brightest target in our campaign, \G, has already been
known as a SNR candidate in the MOST SNR catalogue (Whiteoak
1992). But the limited photon statistics and the poor resolution of
the RASS data do not allow any further probe of its X-ray emission
properties. This has motivated us to observe \G\ with the \emph{Chandra}
X-ray Observatory.  The analysis of this observation is detailed in Hui et al. (2012); in these proceedings, we
present a highlight of the major results.


\section{Confirmation of \G\ as a new SNR}
\begin{figure}
\begin{center}
\includegraphics[width=3.5in]{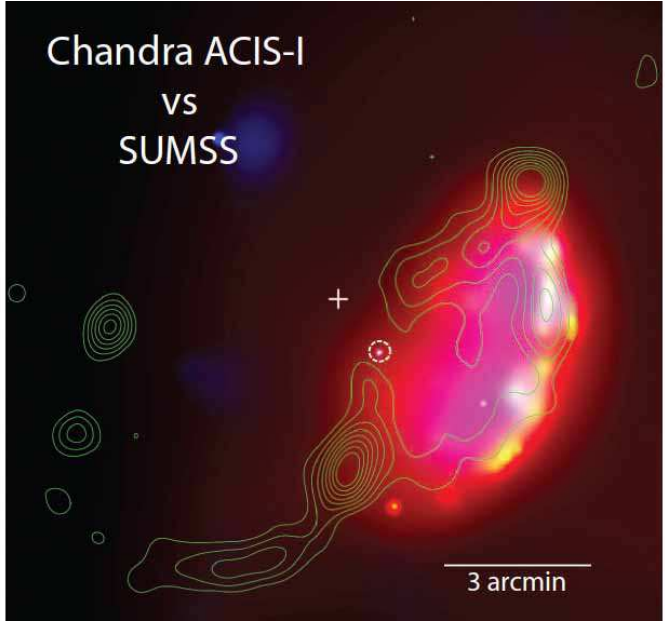}
\caption[]{
$10^{\prime}$$\times$$10^{\prime}$ 
Chandra ACIS-I X-ray colour image of 
\G\ (red: $0.5-1$~keV, green: $1-2$~keV, blue: $2-8$~keV). The 
binning factor of this image is $2^{"}$. Adaptive smoothing has been 
applied to achieve a minimum signal-to-noise ratio of 3. 
The geometrical center inferred from the X-ray morphology is illustrated
by the cross. A bright source which locates closest to the geometrical center is highlighted by the dash circle.
Top is north and left is east.}
\label{rgb}
\end{center}
\end{figure}

The X-ray image of the field around \G\ obtained by \emph{Chandra} is displayed in Figure~\ref{rgb}. 
An incomplete shell-like X-ray structure is found to be well-correlated with the radio shell structure. 
The radio contours are obtained from the 843~MHz Sydney University
Molonglo Sky Survey (Bock et al. 1999). 
Together with the X-ray spectral analysis of the extended emission which suggests it is a shock-heated plasma with a 
temperature in a range of $kT\sim0.6-1$~keV (see Table~2 and Fig.~8 in
Hui et al. 2012), our observation  unambiguously confirms \G\ 
as a new SNR. A recent radio investigation has come the same conclusion, suggesting
\G\ is a young to middle-aged SNR in the early adiabatic phase of evolution (De Horta et al. 2012). 

\section{Discovery of a new central compact object associated with \G}
\begin{figure}
\begin{center}
\includegraphics[width=2.8in,angle=-90]{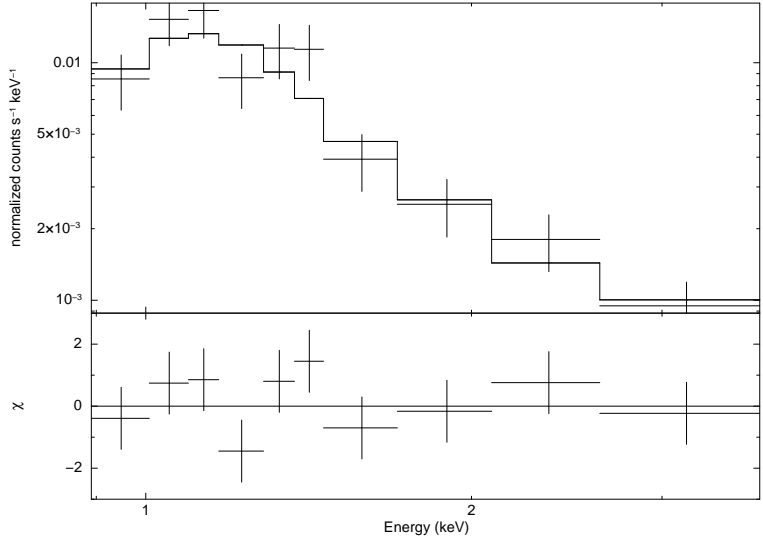}
\caption[]{X-ray spectrum of the emission from the position of CCO as
observed with ACIS-I with the best-fit double blackbody model (\emph{upper panel})
and contributions to the \chisq\ statistics (\emph{lower panel}).
The error bars represent $1\sigma$ uncertainties.}
\label{cco_spec}
\end{center}
\end{figure}

\begin{figure}
\begin{center}
\includegraphics[width=3.2in,angle=90]{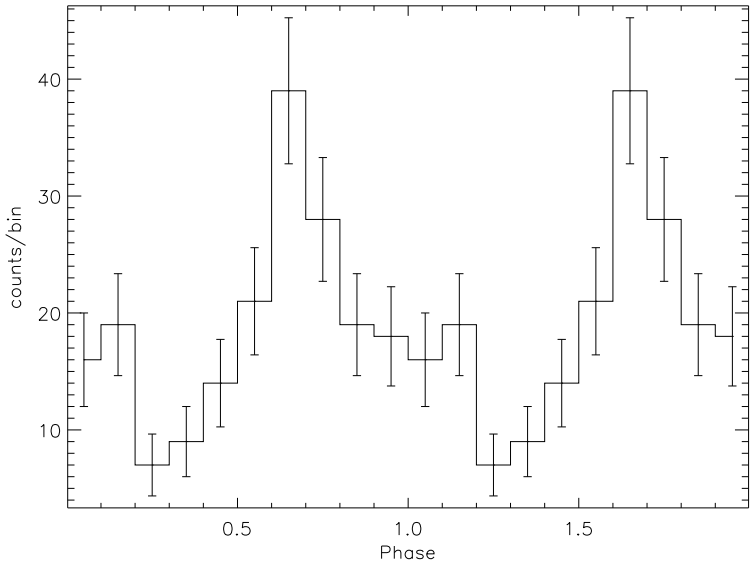}
\caption[]{X-ray counts of CCO versus phase for a periodicity candidate of 
1.4~hrs. Two periodic cycles are shown for clarity. The error bars represent $1\sigma$ uncertainties.}
\label{orbit_fold}
\end{center}
\end{figure}

Apart from confirming the SNR nature of \G, our \emph{Chandra} observation also enables us to search for the possible stellar 
remnant formed in the supernova explosion. Among 17 newly detected X-ray point sources (cf. Table~1 in Hui et al. 2012), 
the brightest source is the one located closest to the geometrical center of \G\ (see Fig.~\ref{rgb}).
Its X-ray point source spectrum can be described by a double blackbody with 
the temperature of $kT_{1}\simeq0.1$~keV, $kT_{2}\simeq0.4$~keV and
emitting areas of $R_{1}\simeq27D_{\rm kpc}$~km and
$R_{2}\simeq35D_{\rm kpc}$~m respectively (see Fig.~\ref{cco_spec}),
where $D_{\rm kpc}$ is the distance to \G\ in units of 1~kpc.
These are similar to those of CCOs -- one of the most enigmatic manifestations of neutron stars (cf. Hui et al. 2006, 2009, 2012). 
The column density inferred from the CCO spectrum is consistent with that for the remnant, which suggest the possible association 
between the CCO and \G.
We proceeded to search for the possible X-ray periodic signals from CCO and have found an interesting periodicity candidate of 
$P\sim1.4$~hrs (Fig.~\ref{orbit_fold}). Together with the spectral energy distribution of its identified optical/IR counterpart,
which conforms with the spectrum of a M dwarf, our results suggest a possible direct evidence for compact binary that survived 
in a supernova explosion (Hui et al. 2012). 

\end{document}